\documentstyle[aps, preprint]{revtex}
\def\beq{\begin{eqnarray}}
\def\eed{\end{eqnarray}}
\begin{document}
\draft
\title{Doping and temperature dependence of incommensurate
antiferromagnetism in underdoped lanthanum cuprates}
\author{Feng Yuan and Shiping Feng}
\address{Department of Physics, Beijing Normal University, Beijing
100875, China\\
Institute of Theoretical Physics, Academia Sinica, Beijing 100080,
China\\
National Laboratory of Superconductivity, Academia Sinica, Beijing
100080, China}

\author{Zhao-Bin Su}
\address{Institute of Theoretical Physics, Academia Sinica, Beijing
100080, China}

\author{Lu Yu}
\address{The Abdus Salam International Centre for Theoretical
Physics, 34014 Trieste, Itlay\\
Institute of Theoretical Physics, Academia Sinica, Beijing 100080,
China}
\maketitle
\date{\today}
\begin{abstract}
The doping, temperature and energy dependence of the dynamical
spin structure factors of the underdoped lanthanum cuprates in
the normal state is studied within the $t$-$J$ model  using the
fermion-spin transformation technique. Incommensurate peaks are
found at $[(1\pm\delta)\pi,\pi]$, $[\pi,(1\pm\delta)\pi]$ at
relatively low temperatures with $\delta$ linearly increasing
with doping at the beginning  and then saturating at higher
dopings. These peaks broaden and weaken in amplitude with
temperature and energy,  in good agreement with experiments. The
theory also predicts a  rotation of  these peaks by $\pi/4$ at
even higher temperatures, being shifted to
$[(1\pm \delta/\sqrt{2})\pi,(1\pm \delta/\sqrt{2})\pi]$.
\end{abstract}


In spite of the tremendous efforts dedicated to the studies of
anomalous properties of high $T_{c}$ superconductors, many
important problems still remain open. Among others, the destruction
of antiferromagnetic long range order (AFLRO)  and appearance of
incommensurate antiferromagnetism (IAF) in doped cuprates is one of
the challenging issues for the theory of strongly correlated
electron systems.  Moreover, the interplay of AF and
superconductivity in these compounds is of fundamental importance
for the high $T_{c}$ theory. Experimentally, by virtue of
systematic studies using NMR and $\mu$SR techniques, particularly
the inelastic neutron scattering, rather detailed information on
dynamical magnetic properties has become available now, awaiting
an adequate theoretical interpretation. It has  been   established
that beyond certain critical doping ($\sim 3 \% $) the
commensurate AFLRO disappears, being replaced by IAF,
characterized by incommensurability parameters $\delta$, i.e., the
AF Bragg peaks are shifted from [$\pi$,$\pi$] to four points
[$\pi(1\pm\delta),\pi$], [$\pi,(1\pm\delta)\pi$] \cite{n1}. For
very low dopings $\delta$ varies almost linearly with concentration
$x$, but saturates at higher dopings. These peaks broaden and
weaken in amplitude as the temperature and energy increase. These
features are fully confirmed by the data on lanthanum cuprates
\cite{n1,n2,n3,n4}, and  have also been found recently on yttrium
cuprates \cite{n5}. Theoretically there is a general consensus that
IAF emerges due to doped charge carriers. Several attempts have
been made to make this argument more precise, including the hole
induced frustration \cite{n6}, stripe formation \cite{n7}, spiral
phase \cite{n8} and Fermi surface nesting \cite{n9}. Based on the
phenomenological ansatz of marginal Fermi liquid behavior
\cite{n10} and tight binding calculation, a detailed fitting of the
experimental data was attempted \cite{n11}. Recently experiments
show stronger singularities of AF fluctuations \cite{n12} than what
is anticipated from the phenomenological models. The proximity to a
quantum critical point \cite{n13} was proposed as an alternative
explanation \cite{n12}. However, to the best of our knowledge, no
systematic calculations have been performed  within the standard
strong correlation models for the dynamical spin structure factors
(DSSF) to  confront the experimental data. Exact diagonalization is
limited by system sizes, while the quantum Monte Carlo technique
faces the negative sign problem for lower temperatures \cite{n14}.
Thus it is rather difficult to   obtain conclusive results.

In this paper, using the fermion-spin theory \cite{n15} which
implements properly the local single occupancy constraint, we
calculate explicitly DSSF for cuprates within the $t$-$J$ model
and reproduce all main features found in experiments
\cite{n2,n3,n4,n12}, including peak position as well as
temperature and energy dependence. Apart from the ratio $t/J$
(taken to be 2.5), there are no other adjustable parameters in the
calculations. Moreover, the theory predicts the magnetic peaks
will be rotated by $\pi/4$ at even higher temperatures, i.e., being
shifted to $[(1\pm\delta/\sqrt{2})\pi,(1\pm\delta/\sqrt{2})\pi]$.
To avoid complications due to bilayers we will focus on the normal
state IAF in lanthanum cuprates.

We start from the $t$-$J$ model on a square lattice,
\begin{eqnarray}
H = -t\sum_{i\hat{\eta}\sigma}C^{\dagger}_{i\sigma}
C_{i+\hat{\eta}\sigma}+{\rm h.c.}-\mu \sum_{i\sigma}
C^{\dagger}_{i\sigma}C_{i\sigma}+J\sum_{i\hat{\eta}}
{\bf S}_{i}\cdot {\bf S}_{i+\hat{\eta}},
\end{eqnarray}
with the local constraint $\sum_{\sigma}C^{\dagger}_{i\sigma}
C_{i\sigma}\leq 1$, where $\hat{\eta}=\pm\hat{x}$, $\pm\hat{y}$,
and ${\bf S}_{i}=C^{\dagger}_{i}{\vec\sigma}C_{i}/2$ are spin
operators with ${\vec \sigma}=(\sigma_{x},\sigma_{y},\sigma_{z})$
as Pauli matrices. The single occupancy local constraint can be
treated {\it properly in analytical form} within the fermion-spin
theory \cite{n15} based on the slave particle approach \cite{n151},
$C_{i\uparrow}=h^{\dagger}_{i}S^{-}_{i}$, $C_{i\downarrow}=
h^{\dagger}_{i}S^{+}_{i}$, where the spinless fermion operator
$h_{i}$ describes the charge (holon) degrees of freedom, while the
pseudospin operator $S_{i}$ describes the spin (spinon) degrees of
freedom. In this representation, the {\it low-energy} Hamiltonian
of the $t$-$J$ model (1) can be rewritten as \cite{n15},
\begin{eqnarray}
H = t\sum_{i\hat{\eta}}h^{\dagger}_{i+\hat{\eta}}h_{i}
(S^{+}_{i}S^{-}_{i+\hat{\eta}}+S^{-}_{i}S^{+}_{i+\hat{\eta}})
+ \mu \sum_{i}h^{\dagger}_{i}h_{i} + J_{{\rm eff}}
\sum_{i\hat{\eta}}({\bf S}_{i}\cdot {\bf S}_{i+\hat{\eta}}),
\end{eqnarray}
with $J_{{\rm eff}}=J[(1-x)^{2}-\phi ^{2}]$, where $x$ is the hole
doping concentration, the holon hopping parameter $\phi=\langle
h^{\dagger}_{i}h_{i+\hat{\eta}}\rangle$, and $S^{+}_{i}$ and
$S^{-}_{i}$ as the pseudospin raising and lowering operators,
respectively. It has been shown \cite{n15} that the constrained
electron operator can be mapped exactly using the fermion-spin
transformation defined with an additional projection operator.
However, this projection operator is cumbersome to handle in the
actual calculations, and we have not presented it explicitly  in
Eq. (2). It has also been shown \cite{n15} that such treatment
leads to errors of the order $x$  in counting the number of spin
states, which is negligible for small dopings. Within this
framework the spin fluctuations only couple to spinons, but the
strong correlation between holons and spinons is included
self-consistently through the holon's parameters entering the
spinon propagator. Therefore both spinons and holons are
responsible for the spin dynamics. The universal behavior of the
momentum-integrated DSSF and susceptibility in the underdoped
regime has been calculated within the fermion-spin theory
\cite{n16} by considering spinon fluctuations around the mean-field
(MF) solution, where the spinon part is treated by the loop
expansion to the second order. Following Ref. \cite{n16}, we obtain
DSSF as,
\begin{eqnarray}
S({\bf k},\omega)&=&-2[1+n_{B}(\omega)]{\rm Im}D({\bf k},\omega)
\nonumber \\
&=&-2[1+n_{B}(\omega)]{B^{2}_{k}{\rm Im}\Sigma_{s}^{(2)}({\bf k},
\omega)\over [\omega^{2}-\omega^{2}_{k}-B_{k}{\rm Re}
\Sigma_{s}^{(2)}({\bf k},\omega)]^{2}+[B_{k}{\rm Im}
\Sigma_{s}^{(2)}({\bf k},\omega)]^{2}},
\end{eqnarray}
where the full spinon Green's function, $D^{-1}({\bf k},\omega)=
D^{(0)-1}({\bf k},\omega)-\Sigma_{s}^{(2)}({\bf k},\omega)$, with
the MF spinon Green's function \cite{n17}, $D^{(0)-1}({\bf k},
\omega)=(\omega^{2}-\omega^{2}_{k})/B_{k}$, while
${\rm Im}\Sigma_{s}^{(2)}({\bf k},\omega)$ and ${\rm Re}
\Sigma_{s}^{(2)}({\bf k},\omega)$ are the imaginary and real parts
of the second order spinon self-energy, respectively, obtained
from the holon bubble,
\begin{eqnarray}
\Sigma_{s}^{(2)}({\bf k},\omega)&=&-\left ({Zt\over N}\right )^{2}
\sum_{pp'}(\gamma_{p'+p+k}+\gamma_{k-p'})^{2}{B_{k+p}\over
2\omega_{k+p}}\left ({F_{1}(k,p,p')\over\omega+\xi_{p+p'}-\xi_{p'}
+\omega_{k+p}} \right .\nonumber \\
&-&\left. {F_{2}(k,p,p')\over \omega+\xi_{p+p'}-\xi_{p'}-
\omega_{k+p}}\right ),
\end{eqnarray}
where $\gamma_{{\bf k}}=(1/Z)\sum_{\hat{\eta}}e^{i{\bf k}\cdot
\hat{\eta}}$, $Z$ is the coordination number, $B_{k}=\Delta
[2\chi_{z}(\epsilon\gamma_{k}-1)+\chi(\gamma_{k}-\epsilon)]$,
$\Delta=2ZJ_{{\rm eff}}$, $\epsilon=1+2t\phi/J_{{\rm eff}}$, $F_{1}
(k,p,p')=n_{F}(\xi_{p+p'})[1-n_{F}(\xi_{p'})]+[1+n_{B}
(\omega_{k+p})][n_{F}(\xi_{p'})-n_{F}(\xi_{p+p'})]$, $F_{2}(k,p,p')
=n_{F}(\xi_{p+p'})[1-n_{F}(\xi_{p'})]-n_{B}(\omega_{k+p})[n_{F}
(\xi_{p'})-n_{F}(\xi_{p+p'})]$, $n_{F}(\xi_{k})$ and $n_{B}
(\omega_{k})$ are the fermion and boson distribution functions,
respectively, the MF holon excitation $\xi_{k}=2Zt\chi\gamma_{k}+
\mu$, and the MF spinon excitation, $\omega^{2}_{k}=\Delta^{2}
(A_{1}\gamma^{2}_{k}+A_{2}\gamma_{k}+A_{3})$ with $A_{1}=\alpha
\epsilon(\chi/2+\epsilon\chi_{z})$, $A_{2}=\epsilon[(1-Z)\alpha
(\epsilon\chi/2+\chi_{z})/Z-\alpha(C_{z}+C/2)-(1-\alpha)/(2Z)]$,
$A_{3}=\alpha(C_{z}+\epsilon^{2}C/2)+(1-\alpha)(1+\epsilon^{2})
/(4Z)-\alpha\epsilon(\chi/2+\epsilon\chi_{z})/Z$, the spinon
correlation functions $\chi=\langle S_{i}^{+}S_{i+\hat{\eta}}^{-}
\rangle$, $\chi_{z}=\langle S_{i}^{z}S_{i+\hat{\eta}}^{z}\rangle$,
$C=(1/Z^{2})\sum_{\hat{\eta}\hat{\eta'}}\langle
S_{i+\hat{\eta}}^{+}S_{i+\hat{\eta'}}^{-}\rangle$, and $C_{z}=
(1/Z^{2})\sum_{\hat{\eta}\hat{\eta'}}\langle S_{i+\hat{\eta}}^{z}
S_{i+\hat{\eta'}}^{z}\rangle$. In order to satisfy the sum rule
for the correlation function $\langle S^{+}_{i}S^{-}_{i}\rangle
=1/2$ in the absence of AFLRO, a decoupling parameter $\alpha$
has been introduced in the MF calculation, which can be regarded
as the vertex correction \cite{n17,n18}. These MF order parameters
$\chi$, $C$, $\chi_z$, $C_z$, $\phi$, and decoupling parameter
$\alpha$ have been determined \cite{n17} by the self-consistent
equations.

Of course, at vanishing dopings the AFLRO gives rise to a
commensurate peak at $[1/2,1/2]$ (hereafter we use the units of
$[2\pi,2\pi]$), which is not presented here  for the sake of space.
Instead, we plot DSSF $S({\bf k},\omega)$ in the ($k_{x},k_{y}$)
plane at doping $x=0.06$, temperature $T=0.1J$ and energy
$\omega=0.05J$ for  $t/J=2.5$ in Fig. 1. The commensurate peak is
split into four IAF peaks at $[(1\pm\delta)/2,1/2]$ and
$[1/2,(1\pm \delta)/2]$. The calculated DSSF  $S({\bf k},\omega)$
has been used to extract the doping dependence of the
incommensurability parameter $\delta(x)$, defined as the deviation
of the peak position from the AF wave vector [1/2, 1/2], and the
result is shown in Fig. 2 in comparison with the experimental data
\cite{n4} taken on La$_{2-x}$Sr$_{x}$CuO$_{4}$ (inset). $\delta(x)$
increases almost linearly with the hole concentration in the
low-doping regime, but it saturates at higher dopings, in full
agreement with experimental data.

For a better understanding of the IAF we have made a series of
scans for $S({\bf k},\omega)$ at different temperatures and
energies, and the result for doping $x=0.06$, $t/J=2.5$ at $T=0.1J$
and $\omega=0.1J$ is shown in Fig. 3. Comparing it with Fig. 1 for
the same set of parameters except for $\omega=0.05J$, we see that
at relatively low temperatures ($T=0.1J$), although the positions
of IAF peaks are almost energy independent, these peaks are
broadened and suppressed with increasing energy, and tend to vanish
at high energies. This reflects that the spin excitations are
rather sharp in momentum space at low temperatures and energies,
then the linewidth, or the  inverse lifetime increases with
increasing energy, in full agreement with experiments
\cite{n4,n12}. Now we turn to discuss the temperature dependence
of $S({\bf k},\omega)$.  $S({\bf k},\omega)$ at $x=0.06$ for
$t/J=2.5$ and $\omega=0.05J$ at temperature $T=0.5J$ is plotted
in Fig. 4. To our big surprise, comparing it with Fig. 1 for the
same set of parameters except for $T=0.1J$, we find that, apart
from the suppression of the peak weight with temperature as
anticipated, the positions of IAF peaks are temperature dependent,
i.e., these peaks deviate from $[(1\pm\delta)/2,1/2]$ and
$[1/2,(1\pm\delta)/2]$ with increasing temperature, and are rotated
by $\pi/4$ in the reciprocal space about $(1/2,1/2)$ at higher
temperatures ($T\geq 0.5J$), being shifted to
$[(1\pm\delta/\sqrt{2})/2,(1\pm \delta/\sqrt{2})/2]$. Up to now
most experimental data show that the positions of IAF peaks in
La$_{2-x}$Sr$_{x}$CuO$_{4}$ \cite{n4,n12} and
La$_{2}$SrCuO$_{4+x}$\cite{n20} are located at
$[(1\pm\delta)/2,1/2]$ and $[1/2,(1\pm\delta)/2]$ in the underdoped
regime, but these data in the normal-state are obtained at
relatively low temperatures (near the superconducting transition).
However, a strong temperature dependence $S({\bf k},\omega)$ in
La$_{2-x}$Sr$_{x}$CuO$_{4}$ has been observed \cite{n4,n12}, namely
the weight of IAF peaks at $[(1\pm\delta)/2,1/2]$ and
$[1/2,(1\pm\delta)/2]$ is suppressed severely with increasing
temperature, whereas the weight is increasing with temperature
\cite{n12} at $[(1\pm\delta/2)/2,(1\pm \delta/2)/2]$. This tendency
is consistent with our theoretical predictions. Experiments at even
higher temperatures are required to check our predictions
explicitly.

Now we give some physical interpretation to the above obtained
results. As seen from Eq. (3), the DSSF has a well-defined
resonance character. $S({\bf k},\omega)$ exhibits a peak when the
incoming neutron energy $\omega$ is equal to the renormalized
spin excitation $E^{2}_{k}=\omega^2_{k}+B_{k}{\rm Re}
\Sigma^{(2)}_{s}(k,E_{k})$, {\it i.e.} $W({\bf k}_{\delta},
\omega)\equiv [\omega^{2}-\omega^{2}_{k_{\delta}}-B_{k_{\delta}}
{\rm Re}\Sigma_{s}^{(2)}({\bf k}_{\delta},\omega)]^{2}=(\omega^{2}
-E^{2}_{k_{\delta}})^{2}\sim 0$ for certain critical wave vectors
${\bf k}_{\delta}$ (positions of IAF peaks). The height of these
peaks is determined by the imaginary part of the spinon
self-energy, {\it i.e.} $1/{\rm Im}\Sigma_{s}^{(2)}
({\bf k}_{\delta},\omega)$. Near half-filling, the spin excitations
are centered around the AF wave vector [1/2, 1/2], so the
commensurate AF peak appears there. Upon doping, the holes disturb
the AF background. Within the fermion-spin framework, as a result
of self-consistent motion of holons and spinons, IAF   is developed
beyond certain critical doping, which means, the low-energy spin
excitations drift away from the AF wave vector, or the zero of
$W({\bf k}_{\delta},\omega)$ is shifted from $[\pi,\pi]$ to
${\bf k}_{\delta}$. As seen from Eq. (3), the physics is dominated
by the spinon self-energy renormalization due to holons. In this
sense, the mobile holes are the key factor leading to IAF. As seen
from Fig. 5, function $W({\bf k},\omega)$ has a rather deep valley
along a circle of radius $\delta$ around [1/2, 1/2]. However, if we
enlarge the scale very significantly (by a factor of 300!!), as
shown in Fig. 6, there is a strong angular dependence with actual
minima (not exactly zero due to precision limitations) at
$[(1-\delta)/2, 1/2]$ and $[1/2, (1-\delta)/2]$ for $T= 0.1J$.
These are exactly the positions of the IAF peaks determined by the
dispersion of very well defined  renormalized spin excitations.
Since the height of the IAF peaks is determined by damping, it is
fully understandable that they are suppressed as the neutron energy
$\omega$ and temperature are increased. The novel result of this
paper, namely the shift of the position for IAF peaks as the
temperature increases is also due to the same reason. To
demonstrate this point in Fig. 6, we plot the function
$W({\bf k},\omega)$ along the arc from ${\bf k}=[(1-\delta)/2,1/2]$
via $[(1-\delta/\sqrt{2})/2,(1-\delta/\sqrt{2})/2]$ to
$[1/2,(1-\delta)/2]$ at doping $x=0.06$ for $t/J=2.5$ and
$\omega=0.05J$ for different temperatures $T=0.1J$ (solid line),
$T=0.2J$ (dashed line), $T=0.4J$ (dash-dotted line), and $T=0.5J$
(dotted line) which shows clearly the temperature dependence of the
minima of  $W({\bf k},\omega)$ -- ${\bf k}_{\delta}$. This means
the self-energy correction due to holon motion is temperature
dependent. From the physical point of view, this is very reasonable.
It would be of great interest to check explicitly this prediction
in neutron experiments at much higher temperatures.

To conclude we have shown very clearly in the paper that if the
local single occupancy constraint is treated properly (as done in
the fermion-spin theory) and the strong spinon-holon interaction
is taken into account, the $t$-$J$ model per se can correctly
reproduce all main features of IAF in underdoped cuprates,
including the doping dependence of the IAF peak position and the
energy as well as temperature dependence of the amplitude of these
peaks, without using adjustable parameters. We believe these are
universal features of the underdoped cuprates, as shown by
experiments on La$_{2-x}$Sr$_{x}$CuO$_{4}$ \cite{n4,n12},
La$_2$CuO$_{4+x}$ \cite{n20} and YBa$_{2}$Cu$_{3}$O$_{7-x}$
\cite{n5}. There might be additional features due to bilayer
splitting in the band structure \cite{n21,n5}, and related
theoretical results will be presented elsewhere \cite{n22}. The
theory also predicts a rotation of IAF peak position at very high
temperatures which should be verified by future experiments.

Finally, we would like to mention  that the influence of the
additional second-neighbor hopping $t'$ on the IAF and
momentum-integrated dynamical spin susceptibility of the $t$-$J$
model has been discussed within the fermion-spin theory. It has
been shown \cite{n23} that for small values of $t'$ the qualitative
behavior of the IAF and integrated dynamical susceptibility of the
$t$-$t'$-$J$ model is the same as  obtained from the present
$t$-$J$ model.

\acknowledgments
The authors would like to thank Professor T. Xiang for the helpful
discussions. This work was supported by the National Natural
Science Foundation under Grant No. 10074007, and the Grant from
Ministry of Education of China.

\begin{figure}
\caption{The dynamical spin structure factor in the $(k_{x},
k_{y})$ plane at doping $x=0.06$ at temperature $T=0.1J$ and
energy $\omega=0.05J$ for parameter $t/J=2.5$.}
\end{figure}

\begin{figure}
\caption{The doping dependence of the incommensurability
$\delta(x)$ of the antiferromagnetic fluctuations. Inset: the
experimental result on La$_{2-x}$Sr$_{x}$CuO$_{4}$ taken from
Ref. [4].}
\end{figure}

\begin{figure}
\caption{The dynamical spin structure factor in the
$(k_{x},k_{y})$ plane at doping $x=0.06$ for parameter
$t/J=2.5$ and energy $\omega=0.1J$ at temperature $T=0.1J$.}
\end{figure}

\begin{figure}
\caption{The dynamical spin structure factor in the
$(k_{x},k_{y})$ plane at doping $x=0.06$ for parameter
$t/J=2.5$ and energy $\omega=0.05J$ at temperature $T=0.5J$.}
\end{figure}

\begin{figure}
\caption{ Function $W({\bf k},\omega)$ in the $(k_{x},k_{y})$
plane at doping $x=0.06$ for parameter $t/J=2.5$ and energy
$\omega=0.1J$ at temperature $T=0.1J$.}
\end{figure}

\begin{figure}
\caption{Function $W({\bf k},\omega)$ from ${\bf k}_{1}=
[(1-\delta)/2,1/2]$ via ${\bf k}_{2}=[(1-\delta/\sqrt{2})/2,
(1-\delta/\sqrt{2})/2]$ to ${\bf k}_{3}=[1/2,(1-\delta)/2]$ at
doping $x=0.06$ for parameter $t/J=2.5$ and energy $\omega=0.05J$
at temperatures $T=0.1J$ (solid line), $T=0.2J$ (dashed line),
$T=0.4J$ (dash-dotted line), and $T=0.5J$ (dotted line).}
\end{figure}

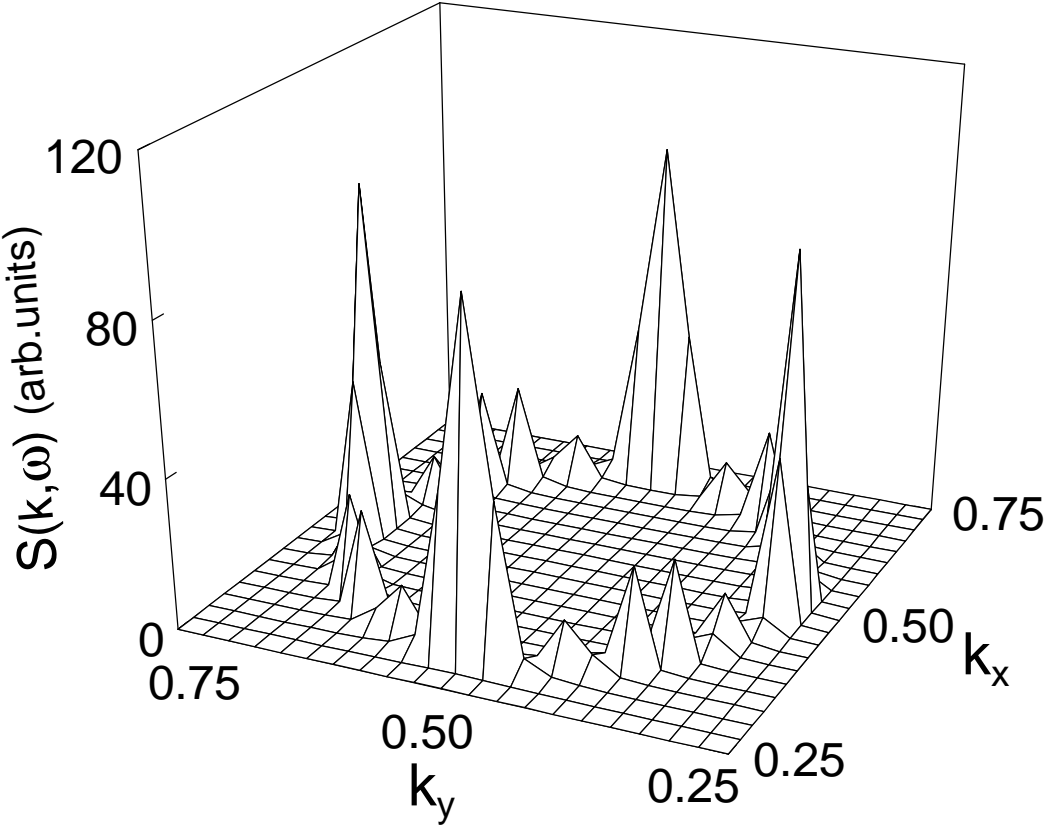
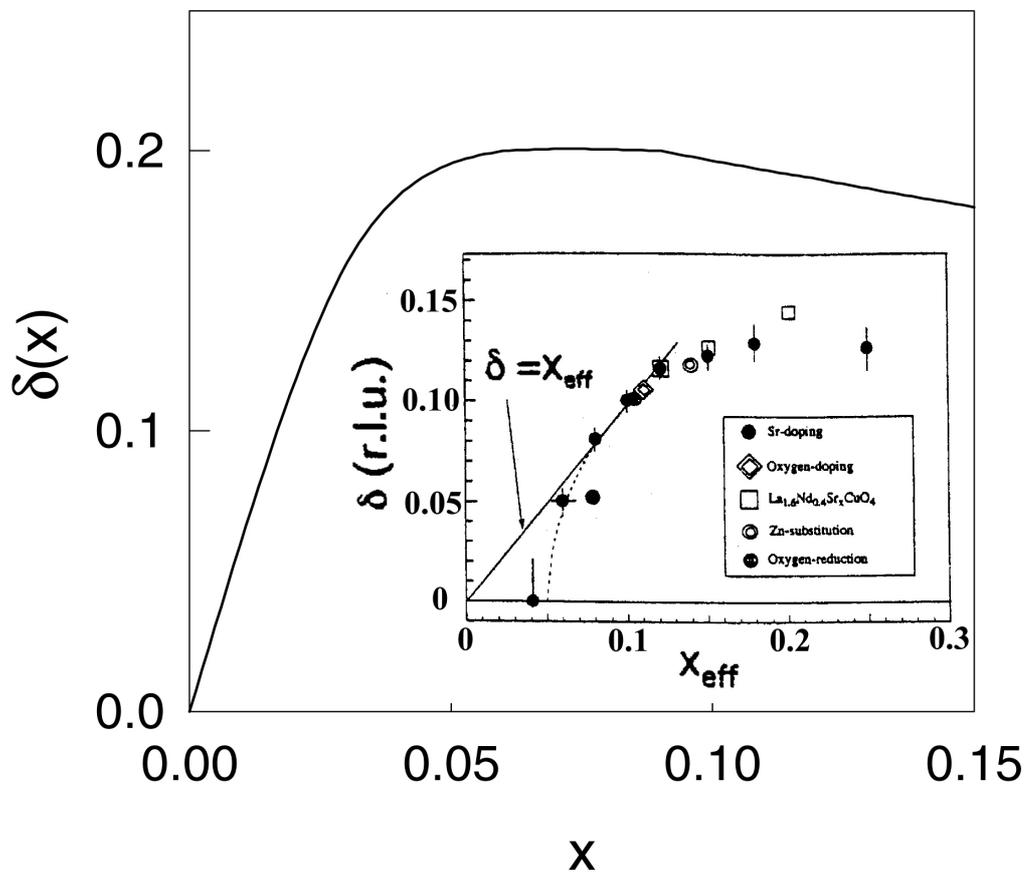
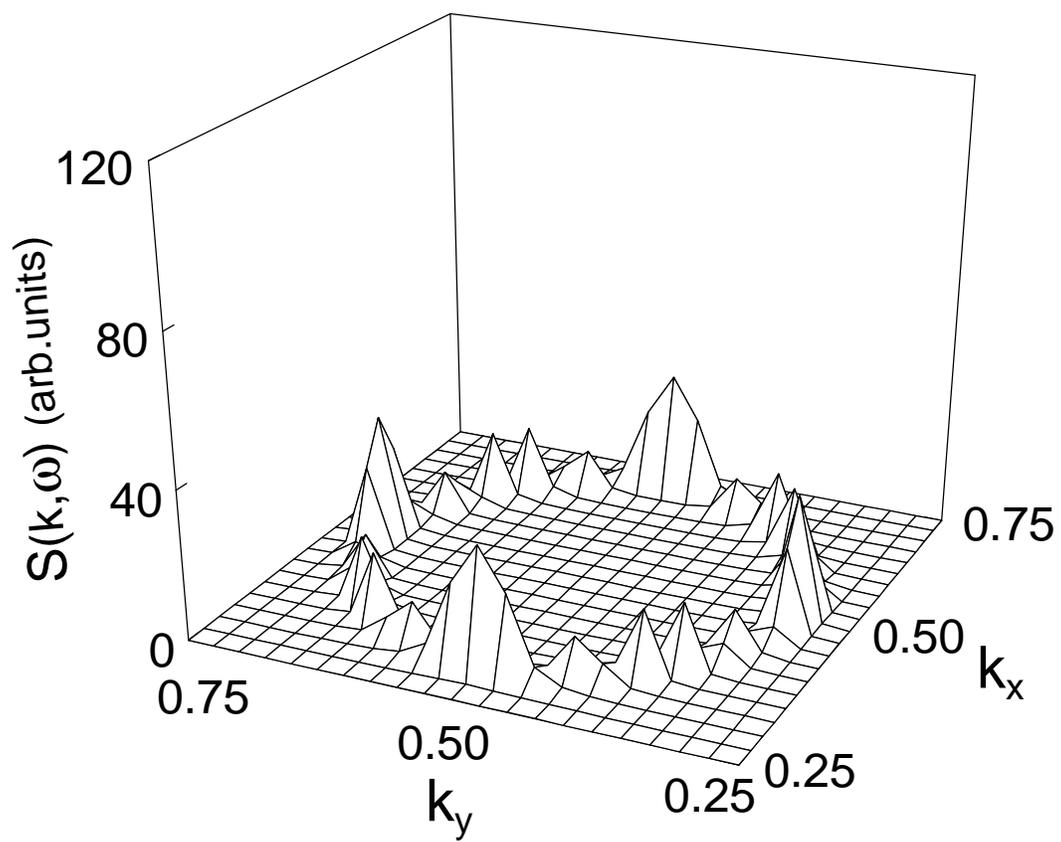
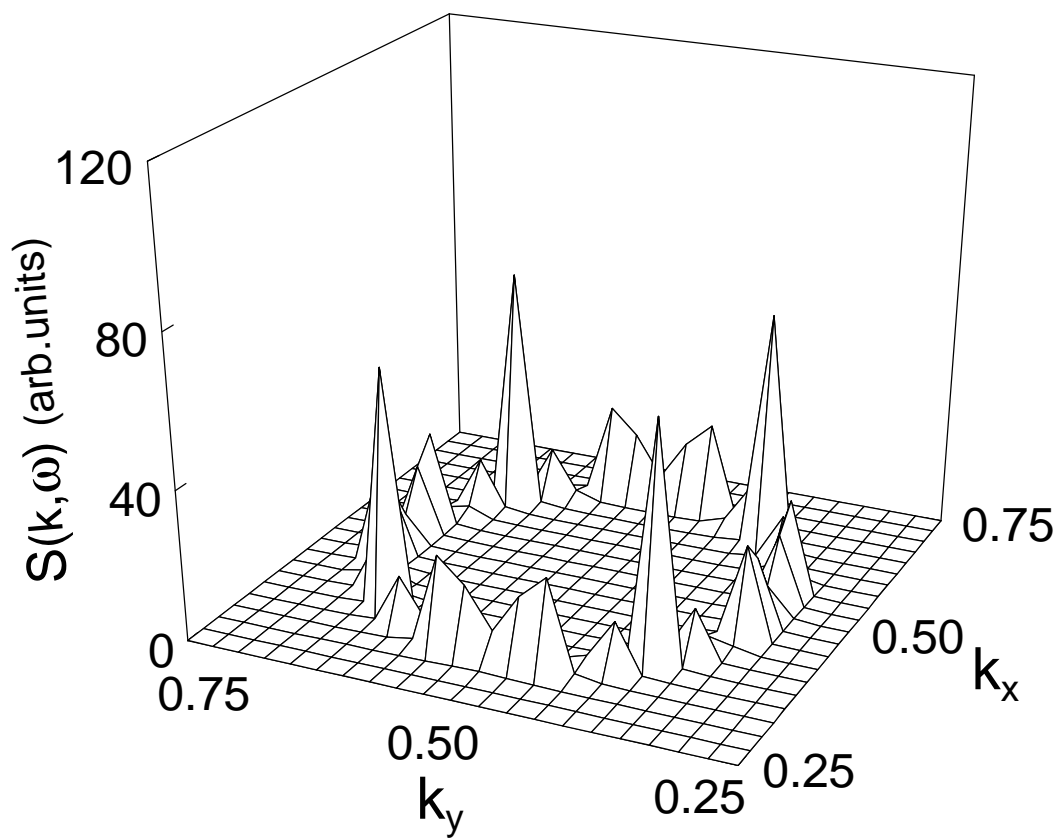
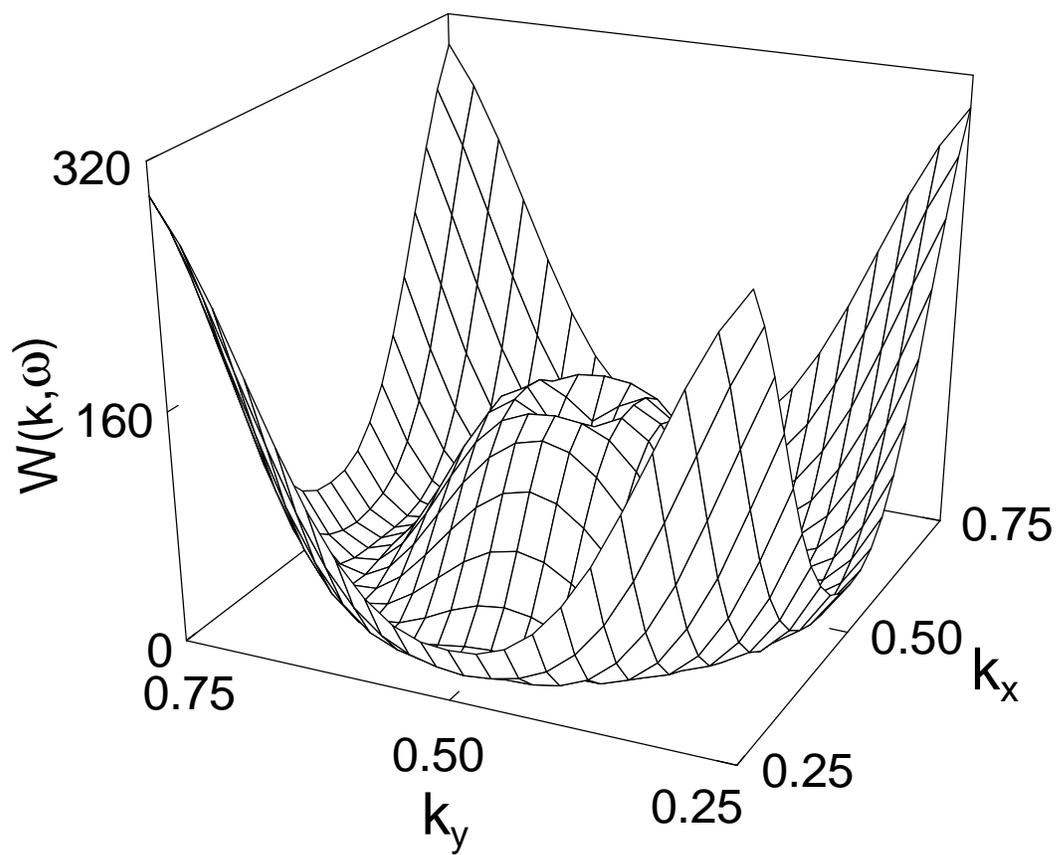
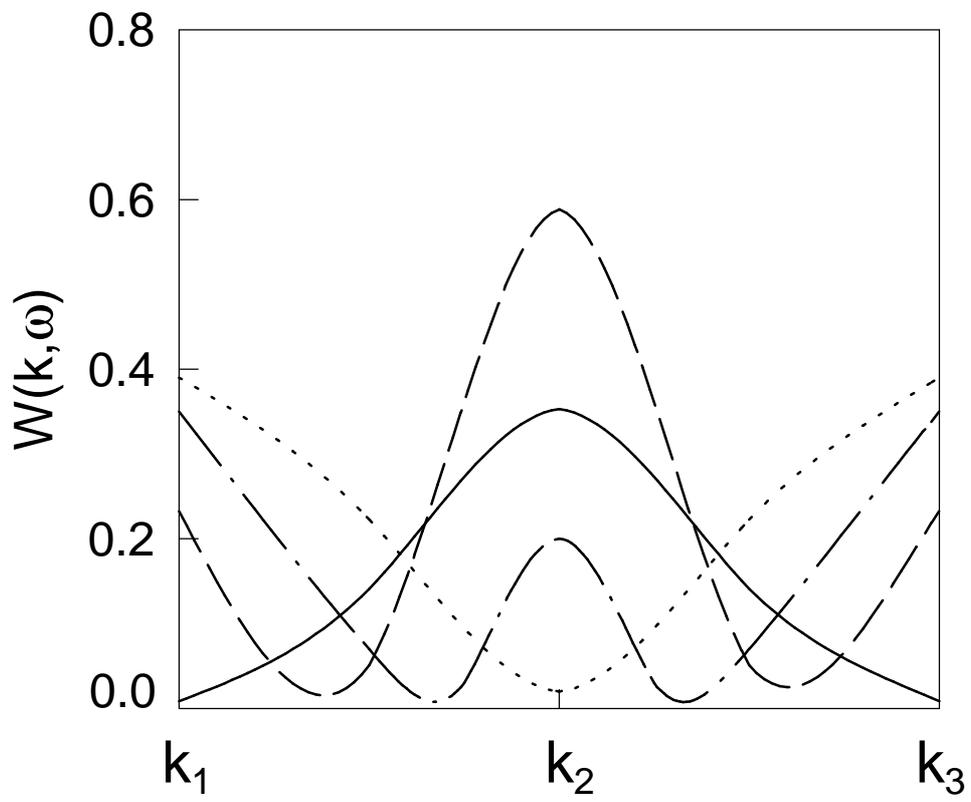
\end{document}